\documentclass{sigchi}

\toappear{}

\usepackage{graphicx}
\usepackage{epstopdf} %necessary for Miktex to read .eps
\usepackage[hyphens]{url} 
\usepackage{tabularx}
\usepackage{paralist}

\newcommand{\spara}[1]{\smallskip\noindent{\bf #1}}
\newcommand{\mpara}[1]{\medskip\noindent{\bf #1}}

\title{Characterizing the Life Cycle of Online News Stories\\Using Social Media Reactions}

\numberofauthors{4}
\author{
\alignauthor Carlos Castillo\\
  \affaddr{Qatar~Computing Research~Institute}\\
  \affaddr{Doha, Qatar}\\
  \email{chato@acm.org}
\alignauthor Mohammed El-Haddad\\
  \affaddr{Al Jazeera}\\
  \affaddr{Doha, Qatar}\\
  \email{mohammed.haddad\\@aljazeera.net}
\alignauthor J{\"urgen} Pfeffer\\
  \affaddr{Carnegie Mellon University}\\
  \affaddr{Pittsburgh, USA}\\
  \email{jpfeffer@cs.cmu.edu}\\
\alignauthor Matt Stempeck\\
  \affaddr{MIT Center for Civic Media}\\
  \affaddr{Cambridge, USA}\\
  \email{stempeck@media.mit.edu}
}

\begin{document}
\maketitle

\begin{abstract}
This paper presents a study of the life cycle of news articles posted online. We describe the interplay between website visitation patterns and social media reactions to news content. We show that we can use this hybrid observation method to characterize distinct classes of articles. We also find that social media reactions can help predict future visitation patterns early and accurately.

We validate our methods using qualitative analysis as well as quantitative analysis on data from a large international news network, for a set of articles generating more than 3,000,000 visits and 200,000 social media reactions. We show that it is possible to model accurately the overall traffic articles will ultimately receive by observing the first ten to twenty minutes of social media reactions. Achieving the same prediction accuracy with visits alone would require to wait for three hours of data. We also describe significant improvements on the accuracy of the early prediction of shelf-life for news stories.
\end{abstract}

;
%\category{H.3.4}{Systems and Software}{User profiles and alert services}

\section{Introduction}

% The topic is important: online news are booming
Traditional newspapers have been in decline in recent years in terms of readership and revenue; in comparison, digital online news have been steadily increasing according to both metrics.\footnote{http://stateofthemedia.org/2012/overview-4/key-findings/}

Recent surveys have shown that about half of the population of the US gets their news online, and about one third goes online every day for news.\footnote{\url{http://www.people-press.org/2012/09/27/section-2-online-and-digital-news-2/}} 

% We are addressing a research gap on this topic
The study of patterns of consumption of online news has attracted considerable attention from the research community for over a decade. This research started with the analysis of access patterns to websites, and has expanded to include topics such as new engagement metrics, personalized news recommendations and summaries, etc. (see Section~\ref{sec:relwork} for an overview).

One line of research looks at consumption and interaction patterns as a single time series and attempts several prediction tasks on it. For example, predicting total comments from early comments~\cite{lee_2010_popularity,tatar_2001_predicting}, total visits from early visits~\cite{kim_2011_temperature}, etc. More recent works incorporate attributes from each specific article (e.g. topic, source, etc.) into the prediction~\cite{bandari2012pulse}.

We adopt a novel approach, in which we integrate different types of interactions of users with an online news article including visits, social media reactions, and search/referrals.
We evaluate our methods on data from Al Jazeera English, a large international news network, deeply characterizing different classes of articles, and predicting their total number of page views and their effective shelf-life (the effective shelf-life of an article is the time span during which it receives most of its visits).

The characterization and prediction of user behavior around news articles is valuable for a news organization, as it allows them (i) to gain a better understanding of how people consume different types of news online; (ii) to deliver more relevant and engaging content in a proactive manner; and (iii) to improve the allocation of resources to developing stories over their life cycle.

\mpara{Our contributions.}
In this paper we present a qualitative and quantitative analysis of the life cycle of online news stories. Our main contributions are the following:
\begin{itemize}
 \item We find that social media reactions can contribute substantially to the understanding of visitation patterns in online news.
 \item We characterize two fundamental classes of news stories: breaking news and in-depth articles, and describe the differences in users' behavior around them.
 \item We describe classes of short-term audience response profiles to news articles in terms of visits and social media reactions (decreasing, steady, increasing, and rebounding).
 \item We improve significantly the accuracy and timeliness of predictive models of total visits and shelf-life of articles, by incorporating social media reactions. 
\end{itemize}

The remainder of this paper is organized as follows. Section~\ref{sec:relwork} provides an overview of previous works related to ours. Section~\ref{sec:framework} introduces our data collection and defines the concepts and variables we use. The main results of our paper are presented as descriptive and predictive analysis in the two following sections: Section~\ref{sec:classification} describes user behavior with respect to different classes of articles, and Section~\ref{sec:prediction} demonstrates the importance of incorporating social media information into the predictive modeling of visits. The last section concludes the paper.

\section{Related work}\label{sec:relwork}

One of the earliest published studies of user behavior in online news was conducted by Aikat~\shortcite{aikat_1998_news}, who studied the web sites of two large newspapers from November 1995 to May 1997. This work describes many of the patterns still seen in news sites today: visits occur mostly during weekdays and working hours; readers ``skim'' pages for information so dwell times tend to be short, and there are clear traffic ``bursts'' that can be attributed to specific news developments. 

With the advent in recent years of what can be considered as new forms of journalism (blogs) and new propagation mechanisms for news (micro-blogs and online social networking sites), the volume of research publications in this area has increased considerably. In this section we overview a few previous works closely related to ours, but our coverage is by no means complete.

\spara{Behavioral-driven article classification.} Previous works including~\cite{crane_2008_response,lehmann_2012_dynamical} that have studied online activities around online resources (e.g. visiting, voting, sharing, etc.), have consistently identified broad classes of temporal patterns. These classes can be generally characterized, first, by the presence or absence of a clear ``peak'' of activity; and second, by the amount of activity before and after the peak.

Crane and Sornette~\shortcite{crane_2008_response} describe classes of visitation patterns to online videos, and present models that are consistent with propagation phenomena in social networks. 
% These models predict four classes of visitation patterns: (i) endogenous sub-critical, where no peak of activity appears, (ii) endogenous critical, where a peak of activity slowly builds and slowly decays, (iii) exogenous sub-critical, where the peak of activity is very sharp, and (iv) exogenous critical, where the peak of activity grows sharply and slowly decays.
Lehmann et al.~\shortcite{lehmann_2012_dynamical} extend these classes by observing that for Twitter ``hashtags'' (user-defined topics) the distributions of activity in different periods (before/during/after) induce distinct clusters of activity that can be interpreted considering the semantics of each hashtag. Romero et al.~\shortcite{romero_2011_differences} describe how manually-assigned classes of hashtags are related to different shapes of the {\em exposure curve}: the probability that a user will propagate some information (``retweet'' in the case of Twitter) after being exposed to the information by a certain number of her neighbors.

Yang and Leskovec~\shortcite{yang_2011_patterns} describe six classes of temporal shapes of attention. Attention is measured in terms of the number of appearances of a given phrase (of a variation of it) corresponding to an event. The patterns describe the distribution of attention over time, as well as the ordering in which different types of media (professional blogs, news agencies, etc.) ``break'' the story.

In general, previous works have established that the evolution of the popularity of different on-line items depends on their class. Figueiredo et al.~\shortcite{figueiredo_2011_tube} describe how YouTube videos that are posted to a ``top'' page on the website, and videos that are making use of professionally produced content, are different from randomly-chosen videos in terms of their visit patterns.

Recently, researchers at URL shortening service Bit.ly~\cite{bitly_2012_halflife} described how an article's half-life (see definition in Section~\ref{sec:framework}) is affected by topics, extending a previous observation than in general there are some topics that are more time-sensitive than others~\cite{gharan_2010_memes}. For instance, business-related articles have on average a longer half-life, while articles related to politics/celebrities/entertainment have an intermediate one. Sports-related articles have in comparison a shorter half-life. Previously, Bit.ly researchers~\cite{bitly_2011_halflife} have shown that this half-life is also affected by the social media platform where the link is first posted (e.g. links  on Facebook were longer-lived than links on Twitter).% half-life depends on where the link is posted: Twitter = 2.8 hr; Facebook = 3.2 hr; Direct (email/IM) = 3.4 hr; YouTube = 7.4 hr. 

%In \cite{figueiredo_2011_tube}: YouTube popularity depends on the type of video. Uses three classes: top (posted in YouTube top list pages), tomb (removed due to copyright violation) or random. They all exhibit different cdf of visits. Using Crane and Sornette methods \cite{crane_2008_response} they use the fraction of visits at the peak date as a characterization factor (high = spam, medium = quality, low = viral); videos in different classes have different fraction of videos in these categories. Also, they have different distribution of referrals.

\begin{table}[t]
\caption{Selected references on predictive modeling of user behavior, sorted by publication year.}
\hskip-0.5em
\scriptsize\begin{tabular}{p{0.79in}p{0.40in}p{1.75in}}\hline
Reference & Collection & Input / Output \\\hline\hline
Tatar et al. \shortcite{tatar_2001_predicting} & 20 \newline Minutes & {\em input}: publication hour, number of comments after a short time, section; {\em output}: total number of comments \\\hline
Brody, Harnad, and Carr \shortcite{brody_2006_citations} & arXiv \mbox{pre-prints} & {\em input}: short-term article downloads; {\em output}: long-term article citations\\\hline
Lee, Moon, and Salamatian \shortcite{lee_2010_popularity} & DPReviews / Myspace & {\em input}: time to first-comment, inter comment arrival stats; {\em output}: time to last comment\\\hline
Lerman and Hogg \shortcite{lerman_2010_news} & Digg & {\em input}: visits; {\em output}: parameters of models that consider examination and promotion patterns\\\hline
Kim, Kim, and Cho \shortcite{kim_2011_temperature} & Blogs & {\em input}: clicks on first 30 minutes; {\em output}: clicks until end of lifetime\\\hline
Yu, Chen, and Kwok \shortcite{yu_2011_predicting} & Facebook pages & {\em input}: content and media type; {\em output}: number of FB likes/shares of each post\\\hline
Lakkaraju and Ajmera \shortcite{lakkaraju_2011_attention} & Facebook pages & {\em input}: text- and other characteristics of the posting and the page; {\em output}: number of FB likes/shares of each post\\\hline
Szabo and Huberman \shortcite{szabo2012predicting} & YouTube and Digg & {\em input}: views (Y), votes (D) in first 10d (Y), 2h (D); {\em output}: total number of views/votes \\\hline
Bandari, Asur, and Huberman \shortcite{bandari2012pulse} & News \newline aggregator & {\em input}: text analysis incl. topics, named entities, subjectivity, etc., source popularity; {\em output}: tweet count\\\hline
Ruan et al. \shortcite{ruan_2012_prediction} & Tweets & {\em input}: topics, past tweets, content features, user features, etc.; {\em output}: tweet count for a given topic\\\hline
Pinto, Almeida, and Gon\c{c}alves \shortcite{pinto_2013_predicting} & YouTube & {\em input}: time series of views in first 7 days; {\em output}: number of views after 30 days \\\hline
Ahmed et al. \shortcite{ahmed_2013_predicting} & YouTube, Digg, Vimeo & {\em input}: views (Y, V)  and votes (D) over time; {\em output}: predict future popularity
\\\hline
\end{tabular}
\label{tbl:relwork-pred}
\end{table}

\smallskip
We deepen and complement previous works on behavioral-driven characterization of online content, by describing the life-cycle of online news articles considering their visitation patterns as well as their social media reactions.

\spara{Prediction of users' activity.} The prediction of the volume of user activities with respect to on-line content items has attracted a considerable amount of research. This is attested by a number of papers, some of which are outlined in Table~\ref{tbl:relwork-pred}. Another active topic that is closely related, but different, is that of predicting real-world variables such as sales or profits using social media signals (e.g. \cite{gruhl_2005_predictive} and many others).

Over the years, the models used to predict user behavior in social media have increased in complexity. For instance, Bandari et al.~\shortcite{bandari2012pulse} and Ruan et al.~\shortcite{ruan_2012_prediction} incorporate into their models features extracted from the content of the articles, such as topics. Yin et al.~\shortcite{yin_2012_straw} study voting behavior over on-line contents and describe a model that considers that users are divided into two populations: a group that follows the majority opinion, and a group that does not. Myers et al.~\shortcite{myers_2012_external} study models that describe user activity in terms of information propagations, including the presence of external influences, e.g. traditional media sources that can reach vast audiences, such as television networks. Huang et al.~\shortcite{huang_2012_predicting} consider an online model of social activities that evolves over time as more information becomes available.

\smallskip
In contrast with previous works, we focus on the dynamic relation between social media reactions and visits over time, and show that both are useful to understand the differences among classes of articles and to predict future visit patterns.

\spara{Analysis of news visits and social media responses.} 
Dezso et al.~\shortcite{dezso_2006_dynamics} analyze the visits to a large news portal in Hungary. One aspect they study which is closely related to our work is the half-life of articles, which is shown to be distributed according to a power-law across a broad range, with a mean of 36 hours. Agarwal et al.~\shortcite{agarwal_2012_multi} study the actions users perform after reading an article, which include printing, commenting, rating, and sharing through e-mail or social media. Their focus is on performing personalized recommendations, but they also uncover that article topics have an effect on the probability of each action, with a division between articles users read privately and articles they share publicly: ``Users tend to share articles that earn them social prestige and credit but they do not mind clicking and reading some salacious news occasionally in private.''

Social media reactions to traditional news media can vary not only in volume but also qualitatively. Hu et al.~\shortcite{hu_2011_event} record tweets during the broadcast of a speech of the US President. They observe that many tweets refer to the speech in general, except for certain topics which are discussed in more detail.

Finally, social media optimization company SocialFlow describes in a whitepaper~\cite{lotan_2011_news} a comparative study of social media responses to several large media outlets: Al Jazeera, BBC News, CNN, The Economist, Fox News and The New York Times. Among other findings, they note that the probability that a user clicks on a tweet is higher for The Economist ($\approx 19\%$) than for Fox News ($\approx 16\%$), Al Jazeera ($\approx 11\%$) or The New York Times ($\approx 4\%$). However, followers of Al Jazeera are almost twice as likely to retweet article links than followers of the other channels.

\smallskip
In contrast with previous works, we consider jointly traffic to the website and social media reactions, as both constitute acts in which users engage with the news content. Additionally, we quantify the richness of Twitter messages over time measuring entropy and counting unique tweets, and show that these variables are key to more accurate predictions of future visits.

\section{Context and dataset}\label{sec:framework}

In this section we provide some context to our research and describe the dataset that will be used on the remainder of the paper.

\subsection{Traditional news and social media} \label{sec:framework-aje}

Our dataset is provided by Al Jazeera English,\footnote{\url{http://www.aljazeera.com/}} a well-established news organization that reaches hundreds of millions of viewers through its TV channel.
Their website
is divided into five major sections: News, In-Depth, Programmes, Sports, and Weather -- plus a collection of blogs, which is outside the scope of this study. Approximately 40 editors/producers work on the areas of News, In-Depth, and Programmes. 

The editors of Al Jazeera English maintain Facebook and Twitter\footnote{Currently Facebook and Twitter are the two most frequent sources of social media referrals to Al Jazeera. Reddit appears in a third place but only through few articles having extremely high visibility.} accounts (we call them ``corporate accounts'' in the rest of the paper) and use them actively to announce their content.
This seems to be a standard practice adopted by all major media organizations in recent years.
Each account
({\tt facebook.com/aljazeera} and {\tt @AJEnglish})
has over 1.5 million followers as of May 2013. Using these accounts, articles in the News section are shared immediately after being posted online. Articles on the In-Depth and Programmes sections are shared throughout the day with the goal of maximizing audience reach across multiple time zones. %Mohammed says announce, Matt says promote

The corporate social media accounts re-share articles at different times of the day, sometimes up to 4 times, on a schedule determined by editors' judgment and designed to increase user engagement. Close attention is paid to the wording of the items posted in social media, including aspects such as their length and the use of hashtags in the case of Twitter. Editors use a variety of online tools to obtain low-latency analytics of traffic and social media, and to decide which hashtags and keywords to use in their postings.

More than half of the visitors to the Al Jazeera English website are from the USA, the United Kingdom, or Canada. According to an online survey taken by Al Jazeera English in 2011 ($n=4,500$), 18\% of respondents said they used Twitter, 42\% Facebook, and 12\% both.

Social media interactions and traffic to the website can complement or substitute each other. Most frequently, they complement each other: people click on the shared content and visit the website. Sometimes, the social media share can be a substitute for a visit to the article, such as when a video can be viewed directly on the social media site, or when the social media content itself delivers enough information to satisfy users without requiring them to click through to the full article.

For instance, the news ``Pakistan's Malala now able to stand in UK'' (19 Oct 2012) generated an unusually large number of shares on Facebook, but comparatively little traffic on the website. At the time, the student-activist was being treated from nearly-fatal wounds received ten days before, and it is likely that users who were following the story just wanted to express their relief or satisfaction at her recovery. %0.4 fb shares per visit

In summary, for Al Jazeera and for most large news organizations, social media is important both because it attracts more visitors to their website than any other external referrer, as well as because it provides more platforms in which to have an audience. Hence, many news organizations adopt an active role in social media in order to increase this positive effect.

\begin{figure}[t]
\centering\includegraphics[angle=-90,width=.9\columnwidth]{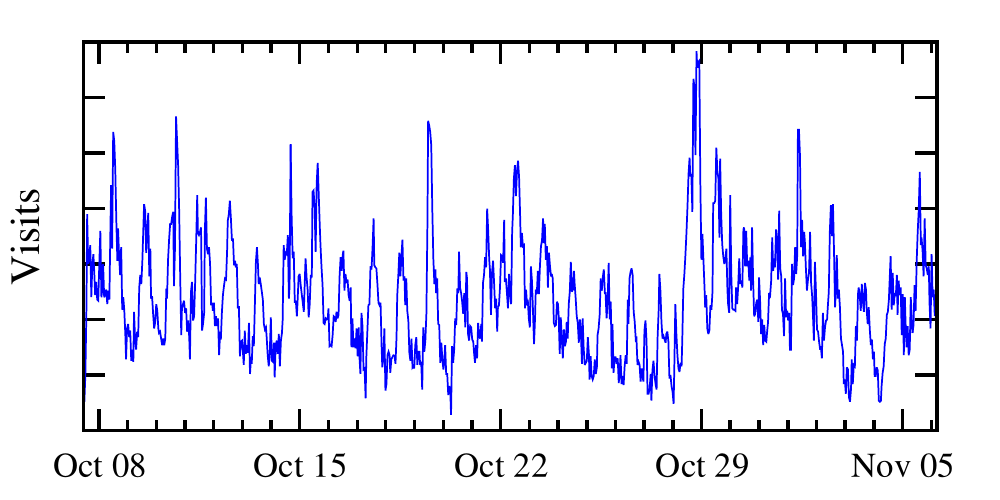}
\caption{Overall visits to articles. Our study considers all articles posted from Oct. 8th to Oct 29th, 2012 (the graph extends until Nov. 6th).}
\label{fig:overall}
\end{figure}

\subsection{Data collection} \label{sec:data-collection}

We focus on a period of three weeks between October 8th, 2012, and October 29th, 2012. The choice of this period is not random: it was a relatively stable period of traffic, only exhibiting a relatively minor peak on October 29th due to Hurricane Sandy. Figure~\ref{fig:overall} depicts the frequency of visits to all the articles in our dataset during the observation period.

The data collection is done via a ``beacon'' embedded in all article pages; this produces events that are processed using Apache S4,\footnote{\url{http://incubator.apache.org/s4/}} a high-performance system for online processing, which is used to collect and aggregate the visits with a 1-minute granularity. For efficiency reasons, only articles obtaining at least 5 visits in a 10-hour window are monitored. The collected data is stored using a Cassandra\footnote{\url{http://cassandra.apache.org/}} NoSQL database.

Our system also collects messages from Facebook (using the Facebook Query Language API) and Twitter (using their Search API). Both platforms have strict limitations on polling frequencies, which impose a trade-off between the number of articles we can monitor and the frequency with which we monitor them. To obtain more accurate results for popular articles, and after experimenting with different settings, we decided to poll social media reactions for articles that are within the list of the 30 most visited articles during each five-minute data collection window. We remark that this list varies considerably over time.

We selected a uniform random sample of articles whose first visit was recorded during the observation period, and kept only those accumulating at least 100 visits during their first week after publication. 
A total of 606 articles was included; this covers over 3.6 million visits and at least 235,000 social media reactions. %This is actually a 100% sample but we don't say so in the paper to avoid giving AJE's competitors an advantage.
% We removed two articles having double URLs
Table~\ref{tbl:overall} presents some summary statistics on this dataset.

\begin{table}[t]
\caption{Summary statistics of our dataset.}
\label{tbl:overall}
\centering
%\begin{tabular}{l|rr}\hline
\scriptsize\begin{tabularx}{1.0\linewidth}{X|rr}\hline
 & Total & Article avg. \\\hline\hline
 Number of articles     & 606   & -\\
 Visits after 1 hour    & 260 K & 430   \\
 Visits after 1 day     & 2.5 M & 4,273 \\
 Visits after 7 days    & 3.6 M & 5,971 \\ \hline
 Facebook shares        & 155 K & 256 \\
 Tweets                 & 80 K  & 133 \\
 Tweet entropy                   & & 5.6 bits \\
 Fraction of unique tweets       & & 19.9 \%  \\
 Fraction of corporate retweets & & 36.8 \%  \\ \hline
 \end{tabularx}
\end{table}

\subsection{Metrics}\label{subsec:variables}

For each article we collected a number of metrics regarding user visits and social media reactions. First, we observed at a granularity of one minute the number of visits (page views) to each article, and the URL of the previous page seen by the users before reaching an article (referral). We bucketed the latter into four classes:

\begin{compactitem}
\item internal links, mostly from the home page of the website: these are the majority of the traffic sources and comprise 70\% of the visits;
\item external links from other sources including social media sites, news aggregators, and others: 14\%;
\item direct links, which have an empty referral and correspond mostly\footnote{\url{http://www.theatlantic.com/technology/archive/2012/10/dark-social-we-have-the-whole-history-of-the-web-wrong/263523/}} to people sharing news through instant messaging, e-mail, or other non-web application: 11\%; and
\item search referrals, basically links from organic search results: 5\%.
\end{compactitem}

We remark that this distribution of referrals corresponds to the articles in our sample, which do not include the homepage of the website, section index pages, or older articles. If we take those into account, the numbers are different, e.g. the search referrals account for 30\% of the visits.

%\smallskip
We also collected periodically the number of times an article has been shared on Facebook, and the content of any Twitter message containing the URL of the article, or a variant of the URL produced by a URL shortening service. We used this data to compute the following variables:

\begin{compactitem}
 \item Number of Facebook shares per minute (interpolated).
 \item Number of tweets per minute.
 \item Number of unique tweets per minute. A tweet is deemed {\em unique} if its edit distance with all previous tweets pointing to the same article (after discarding shortened URLs and ``retweet'' prefixes) is more than 10 characters.
 \item Tweet vocabulary entropy. To compute this, at any given point in time we create a document by concatenating all the tweets received up to that time. Then, we compute the entropy of the distribution of terms in that document.
 \item Number of corporate retweets per minute. A tweet is a ``corporate retweet'' if it includes
 ``RT @AJEnglish'' or ``RT @AJELive''
 in its text. A tweet can be both corporate retweet and unique, as users are free to edit the retweet before posting it.
 \item Number of followers, friends (followees) and statuses of each of the users posting a tweet. %by the end of the observation period
\end{compactitem}

\begin{table}[t]
\caption{Top 10 most frequent words (stemmed and lowercased) in article titles in the ``News'' and ``In-Depth'' sections. Words that appear in both lists are italicized.}
\label{tbl:titles}
\small\centering\begin{tabular}{lcc|lcc}\hline
\multicolumn{3}{c|}{Top words (News, $n=322$)} & \multicolumn{3}{c}{Top words (In-Depth, $n=139$)} \\\hline\hline
Word	&	News	&	In-Depth	&	Word	&	News	&	In-Depth \\\hline
{\em us}	&	34	&	17	&	{\em us}	&	34	&	17	\\
kill	&	21	&	1	&	pictur	&	0	&	10	\\
attack	&	19	&	0	&	obama	&	6	&	6	\\
{\em syria}	&	15	&	4	&	interact	&	0	&	6	\\
dead	&	15	&	1	&	america	&	0	&	6	\\
protest	&	13	&	0	&	muslim	&	0	&	5	\\
rebel	&	12	&	2	&	{\em syrian}	&	11	&	4	\\
vote	&	11	&	3	&	{\em syria}	&	15	&	4	\\
{\em syrian}	&	11	&	4	&	presid	&	4	&	4	\\
pakistan	&	10	&	2	&	polici	&	2	&	4	\\
\end{tabular}
\end{table}

\section{Behavioral-Driven Classes}\label{sec:classification}

In this section we describe classes of articles according to patterns of user behavior.

\subsection{News vs In-Depth}

We observe that articles in the two larger sections of the Al Jazeera English website trigger distinct user behavior patterns: visits and social media reactions on articles in the News section (322 articles in our sample) are different from the ones on articles in the In-Depth section (139 articles). 

\iffalse %CHANGED CAMERA-READY
We depict in Figure~\ref{fig:tagclouds} the most frequent terms in the titles of articles in these two sections -- as in e.g. Lehmann et al.~\shortcite{lehmann_2012_dynamical}.
%
Articles in the News section are dominated by current events, such as the conflict in Syria for this period of time, while articles in the In-Depth section are dominated by photos and analyses of some topics. 

\begin{figure}[t]
\centering\includegraphics[width=.8\columnwidth]{fig/tagcloud_news}
\includegraphics[width=.8\columnwidth]{fig/tagcloud_indepth}
\caption{Most frequent words in article titles in the ``News'' (top) and ``In-Depth'' (bottom) sections.}
\label{fig:tagclouds}
\end{figure}
\fi

\spara{Titles.}
Table~\ref{tbl:titles} includes the most frequent words in titles of articles in these two sections, after converting to lowercase and applying Porter's stemmer.\footnote{\url{http://snowball.tartarus.org/algorithms/porter/stemmer.html}}
While in our sample the US and Syria appear prominently in both sections, articles in the News section include several violent acts, while articles in the In-Depth section are dominated by photos and political analysis. A chi-squared test comparing the entire distributions shows $p<10^{-13}$, rejecting the hypothesis that they are equal.

\begin{figure}[t]
\centering\includegraphics[angle=-90,width=.8\columnwidth]{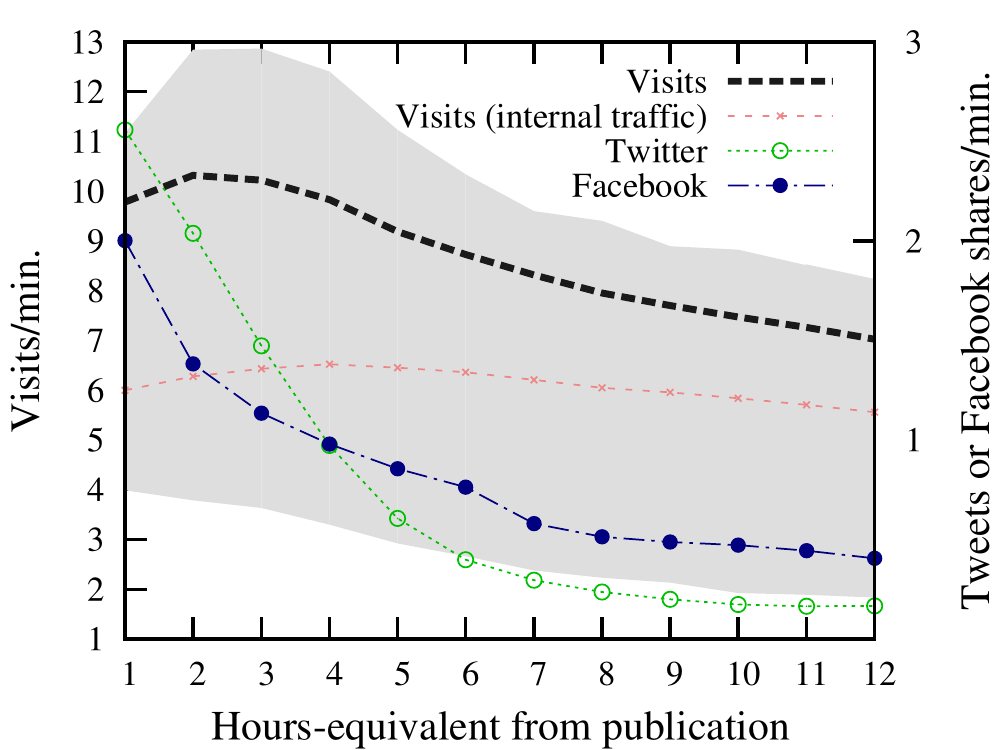}

\includegraphics[angle=-90,width=.8\columnwidth]{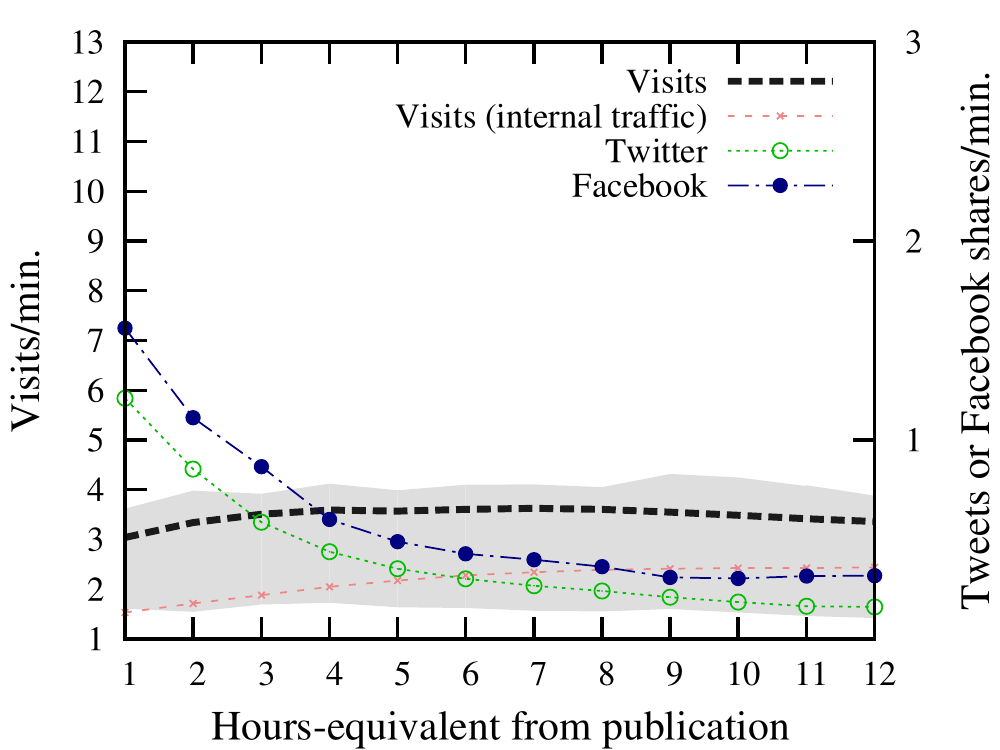}
\caption{Visits per minute (left y-axis) as well as Tweets and Facebook shares per minute (right y-axis) for the first 12 hours. For visits, the shaded area covers 50\% of the data (quantiles 0.25 to 0.75). Top: average  for a News item. Bottom: average  for an In-Depth item.}
\label{fig:series-news}
\end{figure}

\spara{Visits.}
Figure~\ref{fig:series-news}~(top) depicts the average time series of some variables for articles in the News section. Time is expressed in {\em hours-equivalent}, which are hours corrected by the seasonality (day-night, weekday-weekend) of traffic on the website, as in~\cite{szabo2012predicting}. 
Initially there are a number of visits and activity on Twitter and Facebook, that decays rapidly after a short time. This is often the pattern in news media as observed e.g. by~\cite{dezso_2006_dynamics,lotan_2012_aurora} and others. After a few hours, a large amount of visits can be explained by ``internal traffic'', i.e. visitors arriving from the homepage of the site. For most articles, once the news article is displaced from the homepage by more recent items, its traffic slows down considerably.

The profile of visits to In-Depth articles can be more complex. Figure~\ref{fig:series-news}~(bottom) depicts the average series for these articles. We can observe that a sustained level of visits is observed during several hours, as the contents of these articles are not as time-sensitive as those of the News section. We remark that in both cases (News and In-Depth) there is considerable variability from one article to another.

\begin{figure}[t] 
\centering\includegraphics[angle=-90,width=.72\columnwidth]{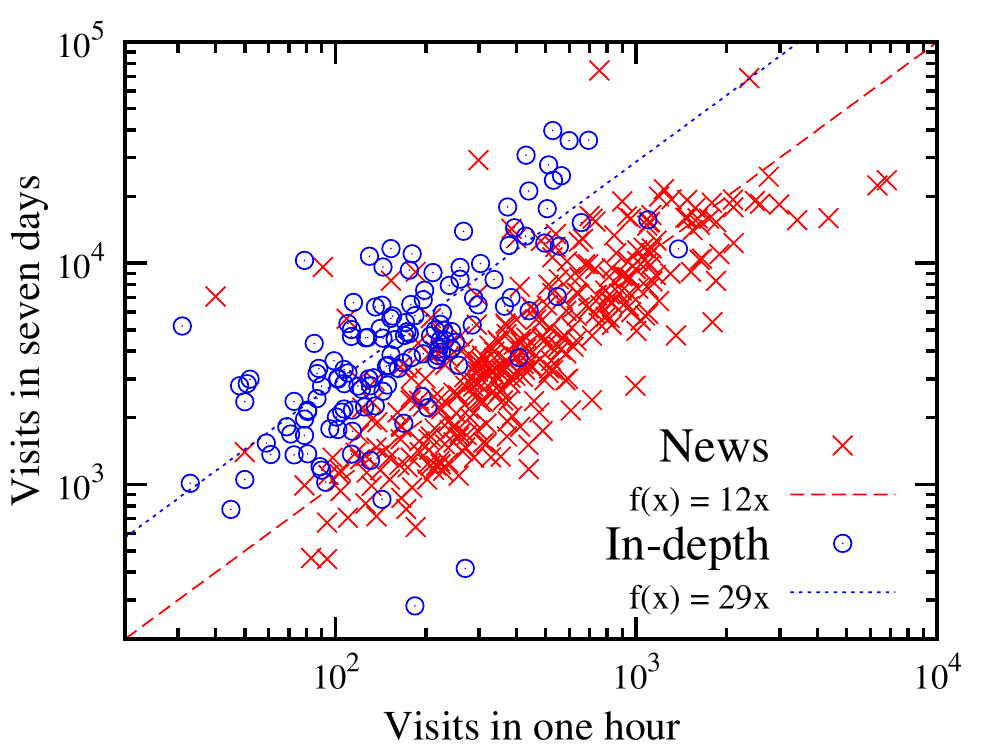}
\caption{Visits in the first hour versus visits on the first week for articles in the two largest categories. A simple function that assumes that visits after seven days are a multiple of visits after one hour has been included, by performing a least-squares fit in the central portion of each distribution.}
\label{fig:scatter_v1h_v7d}
\end{figure}

\begin{figure}[t] 
\centering\includegraphics[angle=-90,width=.72\columnwidth]{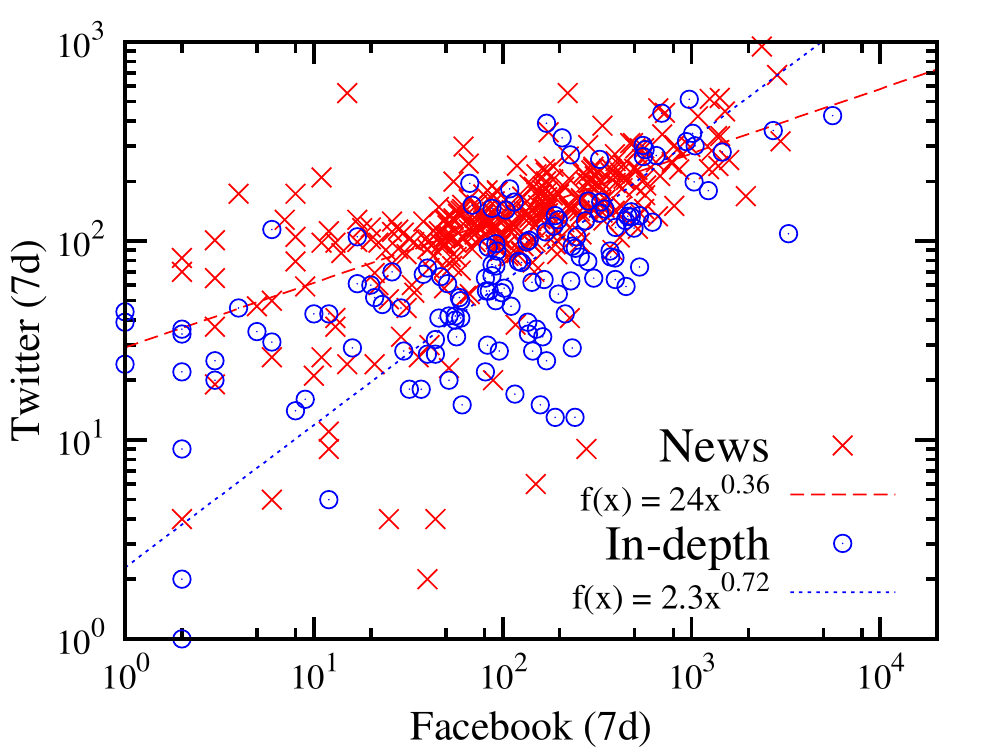}
\caption{Differences in the distribution of Facebook vs Twitter shares. On average the ratio of Facebook shares to tweets is 1.9:1 (1.6:1 for News, 2.7:1 for In-Depth). The result of a least-squares fit in the central portion of each distribution is included.}
\label{fig:scatter_facebook_twitter}
\end{figure}
% cat summary-for-clustering-Clean.csv | awk -F'\t' '{print $4,$31}' | egrep 'indepth' | awk -F' ' '{sum=sum+$2;cnt++}END{print sum/cnt}'

\begin{figure}[t] 
\centering\includegraphics[angle=-90,width=.72\columnwidth]{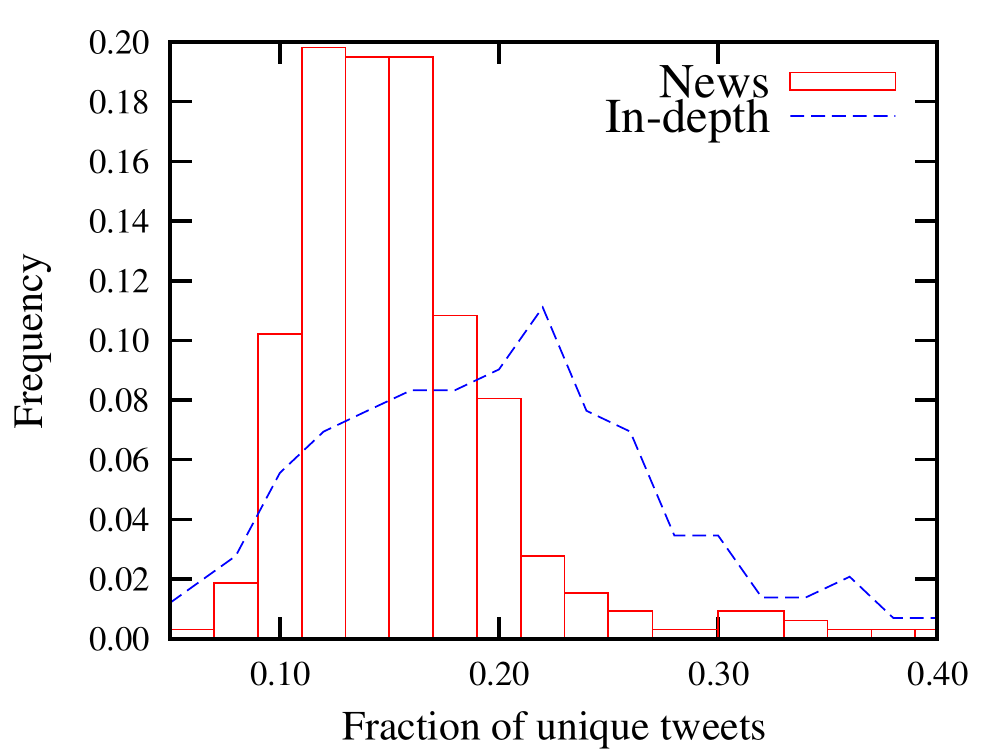}
\caption{Differences in the distribution of the fraction of unique tweets. In both cases, Twitter activity is dominated by re-tweets or repetitions of the same tweets, but In-Depth articles attract more unique tweets.}
\label{fig:hist_tweets_unique_fraction}
\end{figure}

\begin{figure}[t] 
\centering\includegraphics[angle=-90,width=.72\columnwidth]{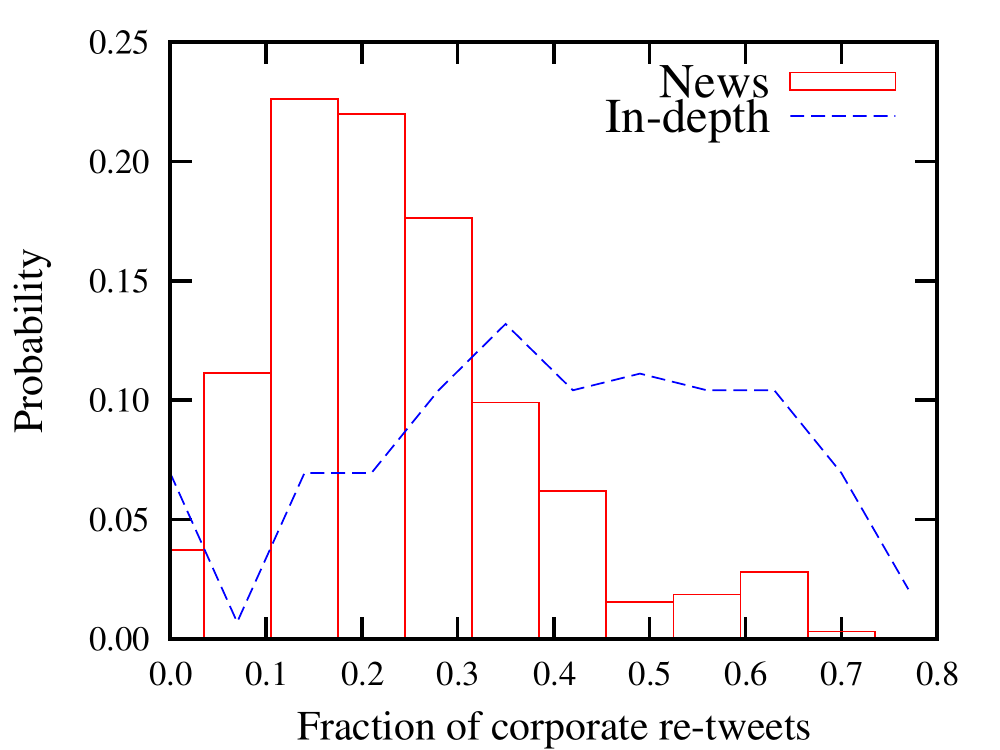}
\caption{Differences in the distribution of fraction of corporate retweets. In-Depth articles have a larger share of re-tweets from the @AJEnglish and @AJELive accounts.}
\label{fig:hist_tweets_corporate_retweets_fraction}
\end{figure}

News items compared to In-Depth items have a more intense first hour, as can be seen in Figure~\ref{fig:scatter_v1h_v7d}. For News, visits in one hour are roughly $1/12$th of the visits in the first week, while for In-Depth they are on average around $1/29$th. The two groups are similar to ``promoted'' (homepage) and ``not-promoted'' stories in Digg as observed in~\cite{szabo2012predicting}.

This difference in behavior can to some extent be explained by the design of the website. News articles are displayed more prominently on the home page, with the most salient location being typically used by a news item; however, In-Depth articles are also visible across the website, including a prominent slot on the top right corner of every page. Additionally to the differences in social media sharing that we discuss next, we observe that long-lived News articles (in terms of effective shelf-life as defined in Section~\ref{sec:shelf-life}) tend to include analysis that would actually make them fit for the In-Depth section. Indeed, the top-3 longer lived News articles in our observation period are ``Profile: Malala Yousafzai'' (Oct 10th, 2012), ``Syrian rebels in uneasy alliances'' (Oct 25th, 2012), and ``Malala is the daughter of Pakistan'' (Oct 13th, 2012); their contents, while motivated by specific news events such as the Syrian conflict and the shooting of a school girl, do not describe the events but rather the context in which they are taking place.

\spara{Social media.}
% cat summary-for-clustering-Clean.csv | awk -F'\t' '{print $4,$29/$28}' | grep indepth | awk -F' ' '{sum=sum+$2;cnt++}END{print sum/cnt}'
On average the ratio of Facebook shares to tweets per article is 1.9:1, which is to some extent consistent with the survey described in Section~\ref{sec:framework-aje} that indicated that there were twice as many website visitors using Facebook as there were Twitter users. Additionally, In-Depth articles are shared more on Facebook given the same level of activity on Twitter, as shown in Figure~\ref{fig:scatter_facebook_twitter}. On average News articles have 1.6 Facebook shares per tweet, while In-Depth articles have~2.7.
% cat summary-for-clustering-Clean.csv | awk -F'\t' '{print $4,$29/$28}' | grep news | awk -F' ' '{sum=sum+$2;cnt++}END{print sum/cnt}'

As shown in Figure~\ref{fig:hist_tweets_unique_fraction} there is also a difference in the number of unique tweets. On average, 17\% of the tweets about News articles are unique, versus 25\% of the tweets about In-Depth articles. This means that a majority of users do not change the content of the tweets when clicking on the ``tweet'' button next to the articles, or when retweeting from another Twitter user.

There is also a difference in the number of corporate retweets, as shown in Figure~\ref{fig:hist_tweets_corporate_retweets_fraction}. On average, 27\% of tweets about News articles are corporate retweets, compared to 44\% of tweets about In-Depth articles. This means that for In-Depth articles a larger share of Twitter activity can be attributed to users who are followers of @AJEnglish or @AJELive, and thus are probably more engaged with these Twitter accounts.

Anecdotally, we know that editors spend more time crafting tweets to promote In-Depth articles than News articles, given that the former are not as time sensitive as the latter. In the case of News, the headline is often posted without modifications to Twitter, which may produce a comparatively less appealing tweet.
%cat summary-for-clustering-Clean.csv | awk -F'\t' '{print $4,$16}' | grep news | awk -F' ' '{sum=sum+$2;cnt++}END{print sum/cnt}'

\subsection{Analysis of news articles} \label{sec:qualitative_news}

We observe that News articles have attention profiles  that are quite predictable, while In-Depth and other article categories show significantly more variability.
We focus on the first 12 hours-equivalent after publication of each article in the News section, and observe the time series of data from all sources including internal links, external links, search engines, and social media (similarly to the time series shown in Figure~\ref{fig:series-news}, but for each individual article instead of as an average). We then classify articles into several classes based on visit patterns that are apparent from these observations, starting with the largest class (``{\em decreasing}'') and following with the other classes. The classification is done by the authors seeking consensus and discussing borderline cases.

At a high level, the classes of articles in our News sample can be roughly described by an ``80:10:10 rule''. The traffic to $\sim$80\% of the articles decreases monotonically during the first 12 hours, the traffic to $\sim$10\% does not decrease, and the traffic to the remaining $\sim$10\% decreases first, but then rebounds. Articles on each  are listed in Appendix~\ref{app:examples}.

Next we provide a brief description of each class and examples of stories appearing in each of them. We remark that in this work we do not attempt to provide a comprehensive content-based typology for news articles within each attention profile.

\mpara{Decreasing (78\%).}
The largest article class represents about 78\% of the sample set. Articles in this class demonstrate an initial spike in visits following article publication, followed by a rather consistent drop in the number of visits, either immediately (244 articles), or after a short delay (7 articles).

Delayed onset traffic decreases have been observed before, such as in~\cite{lotan_2012_aurora} with respect to the shooting in Aurora, Colorado, in 2012. 
This attention pattern can often be attributed to breaking news that resonates with readers located in a time zone that is off-peak when the article is first posted, such as when that portion of the audience is mostly asleep. A story about Hurricane Sandy's movement up the East Coast of the United States, for example, sees an initially sharp visit growth that begins to decline as the East Coast retires for the evening.

The predominance of this class of article indicates that while news itself occurs, and can even be covered, at a constant rate, in most cases readers will only be interested on a news article for a brief period of time after its publication.

\mpara{Steady or Increasing (12\%)}
Roughly 9\% of the sample's articles retain relatively constant visitor rates during their first 12 hours. Compared to news categories with very short shelf-lives, such as sports news, these articles are remarkably consistent. In this subset of news articles, dramatic news and emotional stories appear to garner Facebook shares and, often as a result, extended shelf-lives.

In the U.S., multiple articles on Obama and Romney's sharp-tongued presidential debate drive consistent Facebook and Twitter responses for a relatively long period of time following the articles' publication. A poll on racism in the US has similar staying power and Facebook sharing. In Central Asia, the Taliban attack on Pakistani schoolgirl Malala appears in a number of these articles, where consistent Facebook sharing buoys the article traffic beyond average shelf-life. In Europe, furor over a seismology scandal is posted to Facebook, while in the Middle East, atrocities in the war in Syria and violence between Israel and Hamas also generate hours of steady traffic. Africa sees a new prime minister in Libya, the police shooting of 34 striking miners in South Africa, and a bomb attack on a church in Nigeria, all of which see sustained traffic thanks in part to significant Facebook and Twitter sharing many hours after their initial publication.

Stories in this group were mostly developing stories and many of them had regular updates. One such example is the story about Malala for which Al Jazeera sent a correspondent to the Swat Valley. Being a complex region to cover, a series of news articles and feature stories were written. In addition, Al Jazeera reached out to the Reddit community for a Q\&A session which topped the ``Ask me anything (AmA)'' section\footnote{\url{http://www.reddit.com/r/IAmA/comments/11p6q3/i_am_asad_hashim_journalist_for_the_al_jazeera/}}.
section. 

A relatively small number of articles (3\% of our sample) buck the usual trend and see increased page traffic as time passes after their publication, rather than a decline. To the extent that these articles can be generalized, they resemble the class of articles detailed above. Some of these articles were also updated with supporting content. For at least half of them, web producers added video packages after publication, which may explain to some extent the increase in visits.

\mpara{Rebounding (10\%).}
About 10\% of the articles in our sample initially exhibit a decline in visits/minute, until a point where such decline is reversed. This ``rebound'' occurs either because of internal or external links.

In the case of internal traffic, the traffic patterns behind these rebounding articles sometimes reflect the common newsroom practice of linking to previous coverage in more recent articles. This practice provides additional background context to readers just arriving at the story, but also helps news organizations extract additional value from articles that are otherwise statistically becoming valueless. Stories that required a significant investment of resources to produce are also promoted more heavily than regular articles.  We can see that in these cases, these internal links do indeed deliver readers to articles whose shelf-lives have nearly expired, when measured by homepage and social media traffic. 

The articles that rebound as a result of external traffic are beneficiaries of attention directed from outside of the news organization (e.g. a social networking site, the website of another news network, etc.). Typically each observed burst in external web traffic can be tracked to a single source. Breaking stories can also gain visits as ongoing developments drive significant additional interest. This phenomenon is evidenced, for instance, by three rebounding articles tracking Hurricane Sandy's descent upon the United States.

\bigskip
In general, we see that when News articles cover topics that stray from ``hard news'', the article's attention profile reflects the increased variability seen in the In-Depth pieces. For example, some articles ostensibly cover specific actualities, but also bridge into long-standing issues: in the U.S., ``Immigrant family in pursuit of the American Dream'' and ``Living the modern American Dream'' stoke passions around immigration. %These articles demonstrate more varied fluctuations in visits over time.

The sometimes blurry line between reporting on immediate actualities and longer-term trends like immigration is an area of tension in journalism, one identified by Galtung and Ruge when they asked ``how do 'events' become 'news'''?~\cite{galtung_1965_foreign}.

\section{Improving Traffic Predictions\\Using Social Media Data}\label{sec:prediction}

An increased amount of social media reactions is often correlated with more traffic to online articles. This is particularly marked in the case of non-decreasing and rebounding News articles, as well as In-Depth articles whose visitation patterns are more varied and less predictable than regular (decreasing) News articles. In this section, we combine social media reactions with early visitation measures to provide improved predictions of (i) the volume of visits to an article after 7 days from its publication and (ii) the effective shelf-life of articles, i.e. the time during which they will receive most of their visits. We begin by fitting models to our sample data, and then explore the practicality of this approach for new data.

%\newpage
\subsection{Modeling visiting volume}

Our first goal is to determine to what extent social media reactions can improve the prediction of the overall popularity (total number of visits) of an article. The dependent variable that we want to describe with our models is the total number of visits after 7 days ($v7d$). We use a straightforward approach to answer this question---linear regression models. We include the following variables (described in Section~\ref{subsec:variables}) as observed at the time at which the prediction is performed: number of visits ($v$), number of visits from link referrals ($vr$) and from ``direct'' traffic from e-mail/IM ($vd$), shares on Facebook ($f$), Twitter ($t$), mean number of followers of people sharing on Twitter ($\textrm{\em{foll}}$), entropy of tweets ($ent$), number and fraction of unique tweets ($uni$, $unip$) and fraction of corporate retweets ($cp$).
We use a linear regression model that includes all first-order effects as well as second order interactions. We included second-order interactions because of the interdependency of the variables (e.g. an article with more visits is more likely to have more social media reactions):
%\vskip-1.2em

\begingroup
\setlength{\thinmuskip}{0mu}
\setlength{\medmuskip}{-1mu}
\setlength{\thickmuskip}{-1mu}
\noindent
$lm(vis7d \sim v)$

%\vskip-2.1em
\noindent
$lm(vis7d \sim (v+vr+vd+f+t+\textrm{\em{foll}}+ent+uni+unip+cp)^2)$.
\endgroup

\begin{figure}[t]
\centering\includegraphics[angle=-90,width=.9\columnwidth]{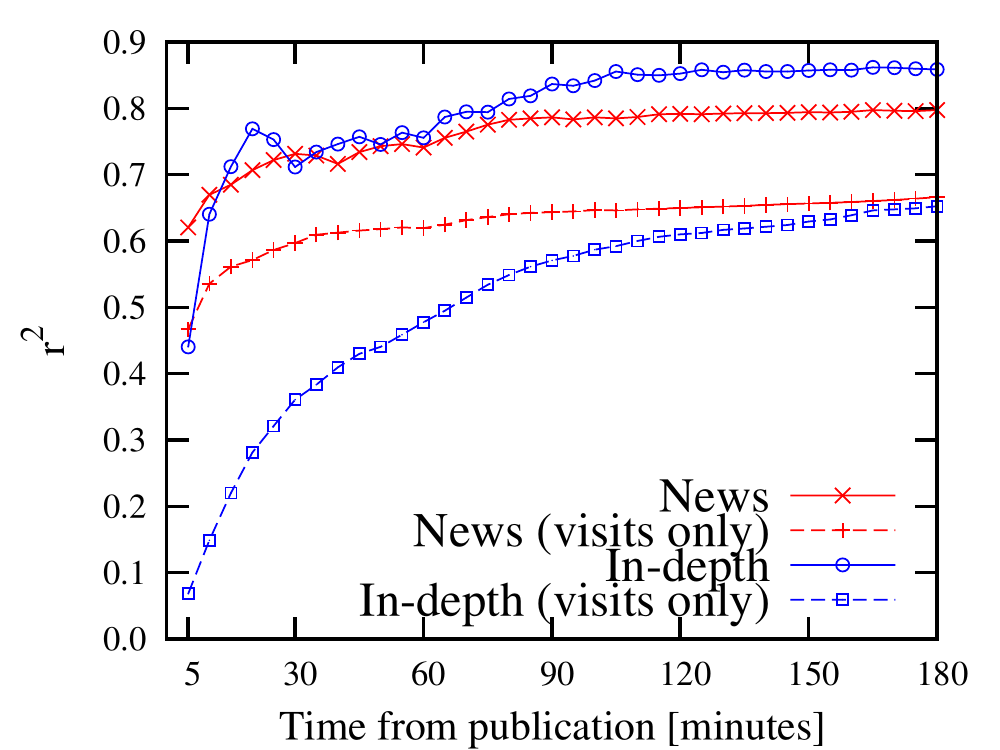}
\caption{Proportion of explained variance ($r^2$) for the prediction of total volume of visits, for News and In-depth articles.}
\label{fig:accuracy_pred_visits}
\end{figure}

\begin{table}[t]
\caption{Modeling visiting volume after 7 days: Significance levels for regression models after 20 minutes.}
\label{tbl:significance}
\setlength{\tabcolsep}{0.5em}
\hskip-0.5em
\small\begin{tabularx}{1.02\linewidth}{Xrlrl}\hline
%\begin{tabular}{lrcrc}\hline
Variable & \multicolumn{2}{c}{In-depth} & \multicolumn{2}{c}{News} \\\hline\hline
Facebook shares & 0.0349 & * & 0.0204 & *  \\
Twitter tweets & 0.0026 & ** & $<$0.0001 & ***  \\
Twitter entropy & $<$0.0001 & *** & 0.0003 & ***  \\
Twitter avg. followers & $<$0.0001 & *** &  &  \\
Volume of unique tweets & & & $<$0.0001 & ***  \\
Unique tweets \% &  & & $<$0.0001 & ***  \\
Corporate retweets \% & 0.0092 & ** &  &  \\
%Traffic from external links \% & 0.2292 & & 0.1615 &  \\
%Traffic from e-mail/IM \% & 0.3892 & & 0.8656 &  \\
\hline
%\end{tabular}
\end{tabularx}
\end{table}

The distribution of visits to articles is log-normal distributed in our data, consistently with previous works~\cite{wu_2007_novelty,bandari2012pulse}. We log-transform ${log (x+1)}$ the visits as well as the volume of social media reactions. For t=5, 10, 15, \dots we calculate the proportion of the explained variance of these two linear models. The result is shown in Figure~\ref{fig:accuracy_pred_visits}.

It takes about 3 hours to be able to explain $>0.6$ of the variance for In-Depth articles, and the additional variables are profitable from the first minutes. After 10-20 minutes we observe the largest difference in our regression models (+0.5 in terms of $r^2$).

We take a closer look at the model variables after 20 minutes to identify the sources of this improvement. For this purpose we stepwise fit the model variables by AIC (Akaike information criterion) as implemented in {\tt stats.step} in R. Table~\ref{tbl:significance} shows the reliability of the Social media variables to serve as good predictor for the volume of visits after 7 days.

The fraction of traffic from different sources does not appear to be a reliable predictor when all variables are used for the model; when we reduce the model to exclusively these two variables, the traffic from e-mail/IM is a more reliable predictor than the traffic from external links.

Social media variables, particularly the number of tweets and the entropy of the vocabulary used in them, seem to be reliable predictors for both In-Depth and News articles. The number of followers of people posting an article on Twitter together with the fraction of corporate retweets seem to be particularly important for In-Depth articles. A possible interpretation is that the response to these articles has a larger component driven by influential accounts and the actions of Al Jazeera editors. In contrast, the number and fraction of unique tweets can be used for the prediction of traffic to News articles. Consequently, a rich online discussion around a breaking within its first minutes is a signal of potentially high and sustained user interest.

\begin{figure}[t]
\centering\includegraphics[angle=-90,width=.9\columnwidth]{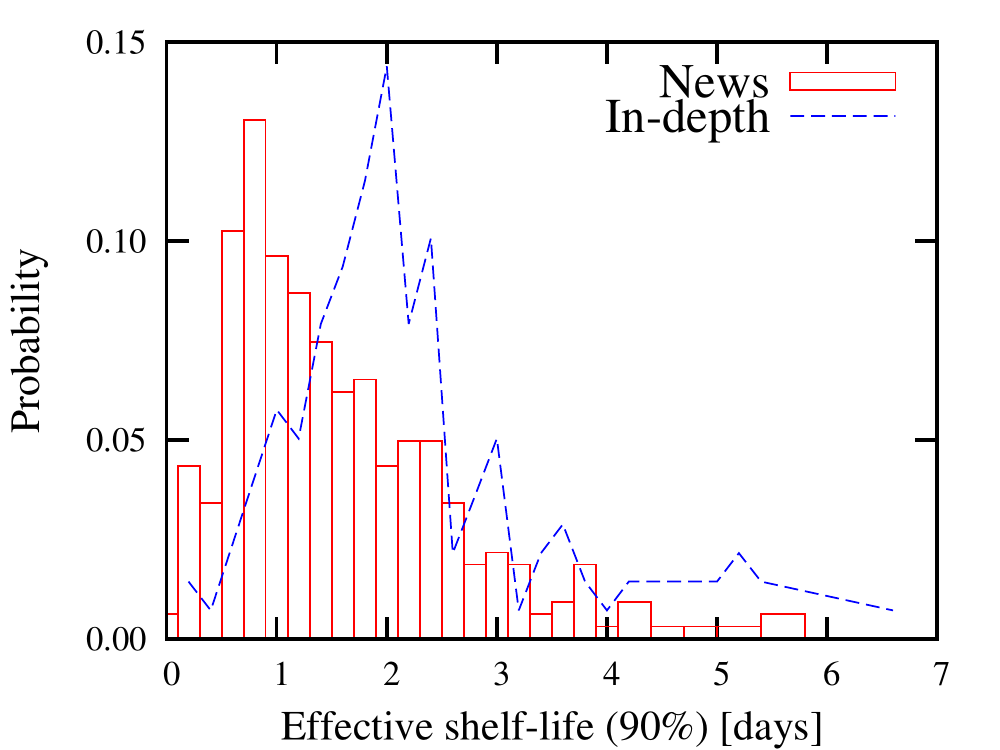}
\caption{Distribution of effective shelf-life.}
\label{fig:hist_shelflife}
\end{figure}

\begin{table}[t]
\caption{Modeling effective shelf-life: Significance levels for regression models after 20 minutes.}
\label{tbl:significance2}
\small
\setlength{\tabcolsep}{0.5em}
\hskip-0.5em
\begin{tabularx}{1.02\linewidth}{Xrlrl}\hline
%\begin{tabular}{lrcrc}\hline
Variable & \multicolumn{2}{c}{In-depth} & \multicolumn{2}{c}{News} \\
\hline\hline
Visits $R^2$ & 0.0005 & & 0.0921 & \\
Social media $R^2$ & 0.4457 & & 0.2193 & \\
Social media $R^2$ adjusted & 0.2274 & & 0.1505 & \\
\hline
Twitter tweets & 0.0138 & * & 0.0061 & ** \\
Twitter entropy & 0.0027 & ** & 0.0024 & ** \\
Twitter avg. followers &  &  & 0.0001 & *** \\
Volume of unique tweets & 0.0026 & ** &  &  \\
Unique tweets \% & 0.0190 & * & 0.0445 & * \\
Corporate retweets & 0.0001 & *** &  &  \\
Traffic from e-mail/IM & 0.0482 & * &  &  \\
\hline
%\end{tabular}
\end{tabularx}
\end{table}

\subsection{Modeling shelf-life} \label{sec:shelf-life}

We define the {\em effective shelf-life} $\tau_{\ell}$ of an article as the time passed between its first visit and the time at which it has received a fraction $\ell$ of the visits it will ever receive. In this work we set $\ell=0.90$, but similar values (e.g. $0.85$, or $0.95$) yield similar results to the ones presented here. When $\ell=0.50$ this is equivalent to half-life~\cite{bitly_2011_halflife,bitly_2012_halflife}.

Given that our observation period is finite, we use a seven-day observation period as a proxy for the total number of visits the articles will ever receive, as for basically all the articles in our sample, there is little activity after 3 or 4 days. This is consistent with the experience of Al Jazeera editors and with observations in previous works (e.g.~\cite{yang_2011_patterns}). We remark however that there are rare cases where an article is ``re-born'' after weeks, for instance when it provides background information for a new development.

% cd ~/data/2012_aljazeera/dataset4/20121125_shelf_life_pred 
% cat shelf-per.txt | grep _news_ | awk '{print $3}' | grep -v mins50 | awk '{sum=sum+$1;cnt=cnt+1;}END{print (sum/cnt)/60}'
% cat shelf-per.txt | grep _indepth_ | awk '{print $3}' | grep -v mins50 | awk '{sum=sum+$1;cnt=cnt+1;}END{print (sum/cnt)/60}'
The distribution of the shelf-life for both classes is depicted in Figure~\ref{fig:hist_shelflife}. As observed in the qualitative analysis, the average shelf-life of In-Depth articles, 2 days and 9 hours, is longer than the one of News articles, 1 day and 16 hours. Their average half-lives are respectively 20 hours and 8 hours (both are shorter than the 36 hours observed by %Dezso et al.~
\cite{dezso_2006_dynamics}). 

\iffalse
%, as shown in Figure~\ref{fig:scatter_visits_shelflife} 
\begin{figure}
\includegraphics[angle=-90,width=\columnwidth]{fig/scatter_visits_shelflife}
\caption{Scatter plot showing the absence of correlation between shelf-life and total visits.}
\label{fig:scatter_visits_shelflife}
\end{figure}
\fi

We observe that the effective shelf-life of all articles is independent from their total number of visits after 7 days (Pearson's correlation $r = -0.03$). This will lead to low accuracy when predicting based solely on visits. For the predictive task the linear regression model setup is analogous to the one used to model visiting volume; this time the dependent variable is $\tau_{90}$. Our focus is again on the variables after 20 minutes.  Running the first regression model (only visits) for this time period reveals differences for News and In-Depth stories (Table~\ref{tbl:significance2}). For News stories, at least 9.2\% of effective shelf-life variance can be described, while visits show no predictive information for In-Depth stories. Including social media variables changes this picture dramatically. Especially for In-Depth stories, a significant part of the variance can now be described. Stepwise fitting of the social media models shows that the number of Facebook shares and the traffic from external links are no reliable predictors for effective shelf-life. In contrast, all Tweet variables reach significant levels. For In-Depth articles corporate retweets and traffic from e-mail/IM also serve as reliable predictors. 

In a nutshell, using social media variables to model effective shelf-life of stories can increase the accuracy of early prediction significantly. This is a very promising result for future research given that we describe the effective shelf-life pattern with data from one single time point without the use of time series or other elaborate models.

%The prediction for News after two hours is within 15 hours of the actual value for the model that uses social media signals (down from about 18 hours). For In-Depth articles, there is a larger improvement, going to about 14 hours from 18 hours. The gains from using a longer observation period are not as relevant as in the case of predicting the volume of visits, as show in in Figure~\ref{fig:accuracy_pred_shelflife}. 

%\begin{figure}[h!]
%\includegraphics[angle=-90,width=\columnwidth]{fig/accuracy_pred_shelflife}
%\caption{Average error of shelf-life predictions.}
%\label{fig:accuracy_pred_shelflife}
%\end{figure}

\subsection{Online predictions} \label{sec:system}

\begin{figure}[t] 
\includegraphics[width=\columnwidth]{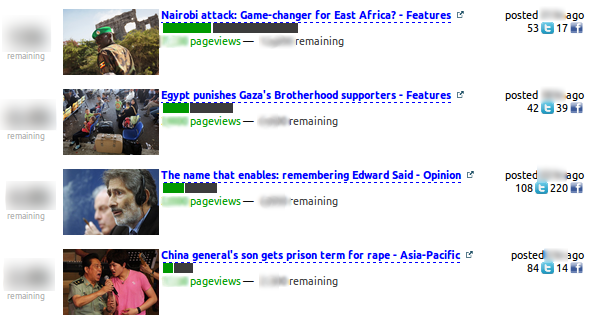}
\caption{Screenshot of http://fast.qcri.org/ depicting predictions for four articles. Green bars indicate number of pageviews so far, gray bars indicate predictions. Exact numbers are business-sensitive so they are omitted.}
\label{fig:fast-screenshot}
\end{figure}

A live system implementing these ideas on data from the Al Jazeera English website is available online.\footnote{\url{http://fast.qcri.org/}} This allows us to further test the effectiveness of our methods in an online setting, in addition to the off-line tests we have described so far. Figure~\ref{fig:fast-screenshot} shows a screenshot of this application.

The system collects data for all articles irrespective of their section, and produces predictions for all articles in the News section using one set of models, and for all the remaining articles (In-Depth, Videos, Programmes, etc.) using another set. In each model set there are models that are executed 1 hour, 6 hours, 12 hours and 24 hours after an article is published. The target variable in this live system is page-views after 3 days. Every 24 hours, it re-trains the models by adding the articles that have passed the 3-day deadline to the training set. After an initial warm up period of 3 weeks, we monitored all 350 article URLs published during a period of 1 week in July 2013 and kept all the predictions done by the system.

First, we evaluated the coverage of our system, which as explained in Section~\ref{sec:data-collection} is designed to focus on the top 30 most visited URLs in every 5-minute period. In practice, we produce predictions within 6 hours for 194 (55\%) of the articles seen. Taking as comparison Google Analytics, which is also used by this website,\footnote{\url{http://analytics.google.com/}} we observe that this covers 65\% of the page-views to Al Jazeera English articles. We remark that this partial coverage is not an intrinsic aspect of the system, but a limitation of using public (instead of paid) access to Twitter's API.

\begin{figure}[t] 
\hskip-1em\includegraphics[angle=-90,width=.5\columnwidth]{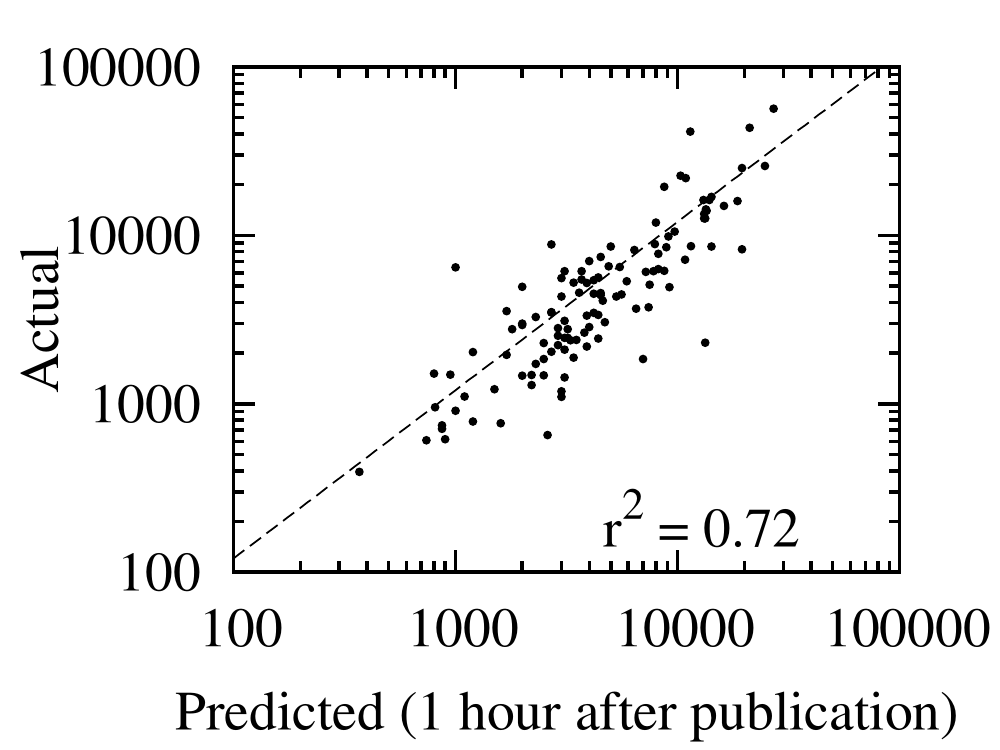}~~\includegraphics[angle=-90,width=.5\columnwidth]{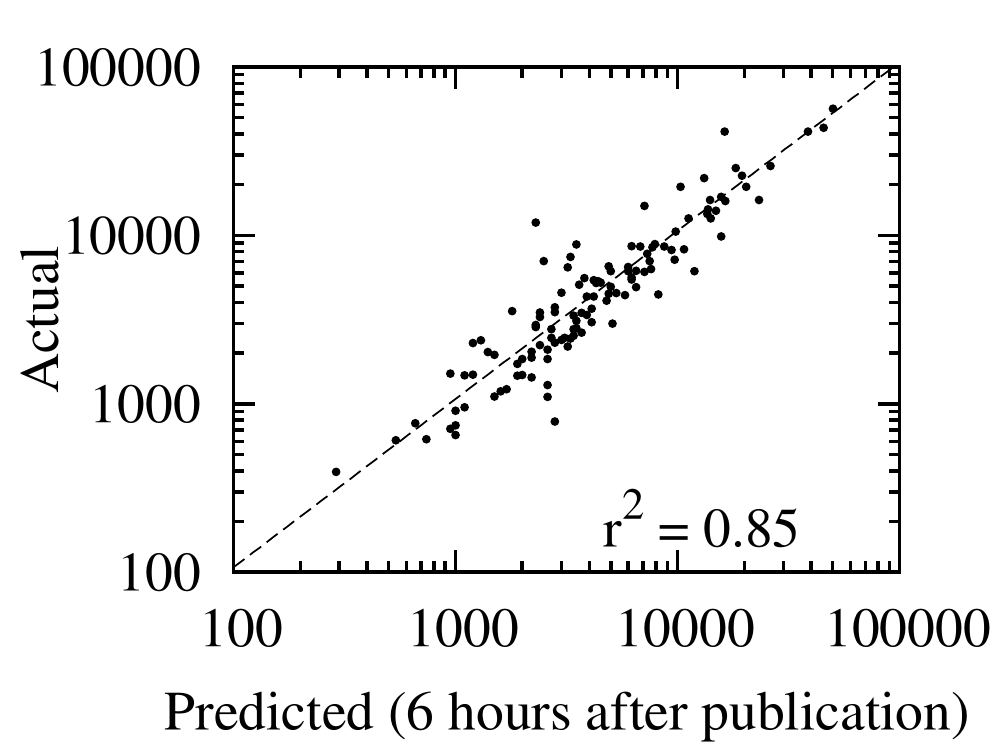}
\caption{Predictions of visits after 3 days using the online system across all articles. Left: predictions 1 hour after publication. Right: predictions 6 hours after publication.}
\label{fig:pred_sys}
\end{figure}

Second, we evaluated the quality of the predictions. In order to do so, we store the predictions done by the different models. Conceptually, each of the 350 articles is in the testing set, and the training set is composed of all the articles published in the period of 1 to 4 weeks before its publication day. Figure~\ref{fig:pred_sys} compares the actual number of page-views with predictions done 1 hour and 6 hours after an article is published. The quality of the predictions after one hour is similar to off-line tests ($r^2=0.72$) even when we are mixing articles from other sections; 134 articles (38\%) not in News or In-Depth -- but we also remark the predictive horizon of the live system is shorter (3 days vs 7 days). Predictions after 6 hours have $r^2=0.85$. The location of the best trade-off between timeliness and accuracy in the range of 1 to 6 hours is an important problem, which requires understanding how editors react to the predictions and use them in practice.

\section{Conclusions}

\mpara{Main findings.}
By adopting an integrated view of users' behavior, we have 
%tracked visits and social media reactions to articles on a top news website over several weeks, and 
observed that there are two classes of articles that generate qualitatively and quantitatively different responses from readers. News articles describing breaking news events tend to decay in attention shortly after they are published and thus have a shorter shelf-life. These articles also have more repetitive social media reactions, as most users simply repeat the news headlines without commenting on them. In-Depth items portraying or analyzing a topic tend to exhibit a longer shelf-life and a richer social media response, including more content-rich tweets in terms of vocabulary entropy and fraction of unique tweets, and more shares on Facebook for the same level of tweets.

By going deeper into the first few hours after publication of News articles, we found three distinctive response patterns in a roughly 80:10:10 proportion: decreasing traffic, steady or increasing traffic, and rebounding traffic. We found that there can be multiple causes for non-decreasing traffic, including the addition of new content to articles, social media reactions, and other types of referrals.

We have shown that social media signals can improve by a large margin the accuracy of predictions of future visits, as well as the accuracy of predictions of article shelf-life. In particular for In-Depth articles which exhibit more complex visit patterns over time, we have found that incorporating measures of the quantity and variety of social media reactions can lead to substantial gains in terms of prediction accuracy.

\mpara{Practical significance.}
From the perspective of a news provider, while no automatic system can replace editorial judgment, understanding and predicting the life cycle of stories has three main benefits:

\begin{itemize}
 \item In the case of News stories, knowing how the audience is interacting with an article is not just ``nice to have'', but increasingly a critical component in delivering timely and relevant content to an ever growing online audience.
 \item For In-Depth stories, which operate on a slower news cycle, knowing when to allocate additional time and resources can significantly improve the news planning process. This is particularly useful for an emerging class of news programmes that combine live online discussions with more traditional TV coverage.
 \item To a web producer, an article with a longer shelf-life means judicious time can be spent preparing backgrounder pieces which are valuable in providing context to a story. From a reach perspective, articles with steady or increasing levels of traffic translate into higher user engagement.
\end{itemize}

Our work depends on having access to a large repository of social media reactions. As more people get into social media (e.g. Twitter), this line of work will become more relevant and will be able to produce even higher quality predictions.

\mpara{Limitations and future work.}
We combine findings from computer science, journalism, and media studies. The research presented here is more difficult to execute than the traditional single-discipline study, but we expect interdisciplinary work on this area to become increasingly common as our media and technology continue to converge.

Our data gathering system collects only aggregate information and does not attempt to link actions across sessions or users across platforms; this prevents us from separating post-read from pre-read sharing, an important distinction explored in~\cite{agarwal_2012_multi}. Another limitation of our work is that we used data from a single website, and we are in the process of gathering data from other sources in order to strengthen our claims. We also used a manual process for categorization of the article classes (decreasing/steady/increasing/rebounding), and we did not attempt a comprehensive content-based classification of articles inside each class.

In this work, we used linear models and did not attempt anything more sophisticated. We do not claim that our models are the more accurate that can be built using this data, but used them to demonstrate in a clear way the importance of social media signals for the predictive tasks we undertake. Better models are definitively possible, and may yield even larger gains in accuracy when incorporating social media signals. 

We also used a data-driven approach in which shelf-life is derived from observations. Alternatively, shelf-life can be derived by fitting a visitation curve produced by a parametrized model~\cite{wu_2007_novelty}. This may lead to an improvement in the prediction accuracy.

\mpara{Reproducibility.}
The data sample used for this study, including feature vectors and the categorization of articles done during the qualitative analysis, is available for research purposes upon request. A~live demo is available at \url{http://fast.qcri.org/}

\mpara{Acknowledgments.}
The authors wish to thank Al Jazeera English for the data used for this study, Kiran Garimella from QCRI for his work in the live system, Janette Lehmann from Yahoo! Research for her valuable help and comments on an early version of this manuscript; Michael K. Martin and Ju-Sung Lee from Carnegie Mellon University for insightful discussions on  regression models; and Edward Schiappa for his feedback on the methodology for the analysis of article classes.

\mpara{Key references:} \cite{brody_2006_citations, szabo2012predicting}

\clearpage
\appendix

\section{Example articles} \label{app:examples}

List of articles in the non-majority classes described in the qualitative assessment of Section~\ref{sec:qualitative_news}.
%All the remaining articles in the News section belong to the ``decreasing'' class.
The data sample is available for research purposes upon request.

\mpara{Delayed decreasing:}
{\scriptsize\begin{verbatim}
Hurricane Sandy moves up US Atlantic coast - Americas
Skydiver lands safely after record jump - Americas
Third-party candidates spar in US debate - Americas
Arrests by French police foiled 'bomb plot' - Europe
Scotland's independence referendum signed - Europe
Rival protesters clash in Egypt's capital - Middle East
Syria opposition 'captures' Assad soldiers - Middle East
\end{verbatim}}

\mpara{Steady:}
{\scriptsize\begin{verbatim}
Bomb attack hits northern Nigerian church - Africa
Libya assembly elects new prime minister - Africa
Police admit 'overreacting' at Marikana - Africa
Marking the Cuban missile crisis - Americas
Obama and Romney face off in final debate - Americas
Obama and Romney meet in combative debate - Americas
Poll finds fresh increase in US racism - Americas
US exports to Iran soar despite sanctions - Americas
Asad Hashim: Ask Me Anything on Malala - Central & South Asia
Clerics declare Malala shooting 'un-Islamic' - Central & South Asia
India suspends Kingfisher licence - Central & South Asia
Pakistani schoolgirl Malala arrives to UK - Central & South Asia
Profile: Malala Yousafzai - Central & South Asia
Teenage rights activist shot in Pakistan - Central & South Asia
Italian seismologists could face jail term - Europe
Karadzic to begin Srebrenica defence at Hague - Europe
Russia says fighters killed in North Caucasus - Europe
Scientists found guilty in Italy quake trial - Europe
Bomb blast hits Damascus' Old City - Middle East
Fatah claims victory in West Bank poll - Middle East
Fighting dims hopes for Syria Eid truce - Middle East
Hariri calls on Lebanese to attend funeral - Middle East
Israel strikes Gaza after Hamas retaliation - Middle East
Marginalisation of disabled people in Egypt - Middle East
Palestinians vote in municipal elections - Middle East
Rights group says Syria used cluster bombs - Middle East
Syrian children killed in Idlib air raids - Middle East
US and EU urge political stability in Lebanon - Middle East
\end{verbatim}} 

\vfill\eject

\mpara{Increasing:}
{\scriptsize\begin{verbatim}
Colombia and FARC rebels launch negotiations - Americas
Immigrant family in pursuit of American Dream - Americas
Living the modern 'American Dream' - Americas
Man charged over attempted US bank bomb plot - Americas
Minors flee Central American violence - Americas
Anti-austerity protests erupt in Athens - Europe
Lithuanians vote out austerity government - Europe
Scientists await verdict in Italy quake trial - Europe
Assault on Yemen base blamed on al-Qaeda - Middle East
Qatari emir in historic Gaza visit - Middle East
\end{verbatim}}

\mpara{Rebounding:}
{\scriptsize\begin{verbatim}
African and EU leaders to hold Mali summit - Africa
Evidence of mass murder after Gaddafi's death - Africa
Nigerian soldiers kill dozens of civilians - Africa
State-linked Libyan militias shell Bani Walid - Africa
Tunisia clash leaves opposition official dead - Africa
UN urges military action plan for Mali - Africa
Wounded Mauritania president flown to Paris - Africa
Argentine crew to vacate ship seized in Ghana - Americas
Armstrong 'unaffected' by doping report - Americas
Biden and Ryan set for crucial VP debate - Americas
Brazil forces set for raid on Rio slums - Americas
Candidates spar in US vice president debate - Americas
Cuba's Castro appears in public - Americas
First planet with four suns discovered - Americas
Forecasters predict 'serious' Hurricane Sandy - Americas
Hurricane Sandy approaches eastern US - Americas
Tsunami warning for Hawaii lifted - Americas
US deficit tops $1 trillion for fourth year - Americas
US East Coast prepares for Hurricane Sandy - Americas
Dozens dead in Afghanistan Eid suicide blast - Central & South Asia
Pakistan court probes bartering of girls - Central & South Asia
Pakistan teen activist in critical condition - Central & South Asia
Berlusconi vows to remain in political arena - Europe
Boxer a big hit as Ukraine readies for vote - Europe
EU leaders agree on banking supervisor - Europe
Germany's Merkel reassures Greece - Europe
Merkel arrives in Greece amid tight security - Europe
Russia demands Turkey explain intercepted jet - Europe
Russian opposition aide arrested - Europe
Baghdad area hit by more deadly Eid attacks - Middle East
Eid truce awaits Syrian government response - Middle East
Kuwait police fire tear gas at protesters - Middle East
Syrian forces continue to shell Aleppo - Middle East
\end{verbatim}}


\begin{thebibliography}{10}

\bibitem{agarwal_2012_multi}
Agarwal, D., Chen, B.~C., and Wang, X.
\newblock Multi-faceted ranking of news articles using post-read actions.
\newblock In {\em Proc. of CIKM}, ACM (2012), 694--703.

\bibitem{ahmed_2013_predicting}
Ahmed, M., Spagna, S., Huici, F., and Niccolini, S.
\newblock A peek into the future: predicting the evolution of popularity in
  user generated content.
\newblock In {\em Proceedings of the sixth {ACM} international conference on
  Web search and data mining} (New York, {NY}, {USA}, 2013), 607--616.

\bibitem{aikat_1998_news}
Aikat, D.
\newblock News on the web.
\newblock {\em Convergence: The International Journal of Research into New
  Media Technologies 4}, 4 (Dec. 1998), 94--110.

\bibitem{bandari2012pulse}
Bandari, R., Asur, S., and Huberman, B.~A.
\newblock The pulse of news in social media: Forecasting popularity.
\newblock Feb. 2012.

\bibitem{bitly_2011_halflife}
{Bitly Science Team}.
\newblock You just shared a link. how long will people pay attention?
\newblock The {Bit.ly} blog, Sept. 2011.

\bibitem{bitly_2012_halflife}
{Bitly Science Team}.
\newblock Halflife by topic.
\newblock The {Bit.ly} blog, Nov. 2012.

\bibitem{brody_2006_citations}
Brody, T., Harnad, S., and Carr, L.
\newblock Earlier web usage statistics as predictors of later citation impact:
  Research articles.
\newblock {\em J. Am. Soc. Inf. Sci. Technol. 57}, 8 (June 2006), 1060--1072.

\bibitem{crane_2008_response}
Crane, R., and Sornette, D.
\newblock Robust dynamic classes revealed by measuring the response function of
  a social system.
\newblock {\em PNAS 105}, 41 (October 2008), 15649--15653.

\bibitem{dezso_2006_dynamics}
Dezs\"{o}, Z., Almaas, E., Luk\'{a}cs, A., R\'{a}cz, B., Szakad\'{a}t, I., and
  Barab\'{a}si, A.~L.
\newblock Dynamics of information access on the web.
\newblock {\em Physical Review E (Statistical, Nonlinear, and Soft Matter
  Physics) 73}, 6 (2006), 066132+.

\bibitem{figueiredo_2011_tube}
Figueiredo, F., Benevenuto, F., and Almeida, J.~M.
\newblock The tube over time: characterizing popularity growth of youtube
  videos.
\newblock In {\em Proc. of WSDM}, ACM (2011), 745--754.

\bibitem{galtung_1965_foreign}
Galtung, J., and Ruge, M.~H.
\newblock The structure of foreign news.
\newblock {\em Journal of Peace Research 2}, 1 (1965), 64--91.

\bibitem{gharan_2010_memes}
Gharan, S.~O., Ronaghi, F., and Wang, Y.
\newblock What memes say about the news cycle?
\newblock Tech. rep., Stanford University, 2010.

\bibitem{gruhl_2005_predictive}
Gruhl, D., Guha, R., Kumar, R., Novak, J., and Tomkins, A.
\newblock The predictive power of online chatter.
\newblock In {\em Proc. of KDD}, ACM (2005), 78--87.

\bibitem{hu_2011_event}
Hu, Y., John, A., and Seligmann, D.~D.
\newblock Event analytics via social media.
\newblock In {\em Proc. of SBNMA}, ACM (2011), 39--44.

\bibitem{huang_2012_predicting}
Huang, S., Chen, M., Luo, B., and Lee, D.
\newblock Predicting aggregate social activities using {Continuous-Time}
  stochastic process.
\newblock In {\em Proc. of CIKM 2012} (2012).

\bibitem{kim_2011_temperature}
Kim, S.-D., Kim, S.-H., and Cho, H.-G.
\newblock Predicting the virtual temperature of {Web-Blog} articles as a
  measurement tool for online popularity.
\newblock In {\em Computer and Information Technology (CIT), 2011 IEEE 11th
  International Conference on}, IEEE (Aug. 2011), 449--454.

\bibitem{lakkaraju_2011_attention}
Lakkaraju, H., and Ajmera, J.
\newblock Attention prediction on social media brand pages.
\newblock In {\em Proc. of CIKM}, ACM (2011), 2157--2160.

\bibitem{lee_2010_popularity}
Lee, J.~G., Moon, S., and Salamatian, K.
\newblock An approach to model and predict the popularity of online contents
  with explanatory factors.
\newblock In {\em IEEE Conference on Web Intelligence} (2010).

\bibitem{lehmann_2012_dynamical}
Lehmann, J., Gon\c{c}alves, B., Ramasco, J.~J., and Cattuto, C.
\newblock Dynamical classes of collective attention in twitter.
\newblock In {\em Proc. of WWW}, ACM (2012), 251--260.

\bibitem{lerman_2010_news}
Lerman, K., and Hogg, T.
\newblock Using a model of social dynamics to predict popularity of news.
\newblock In {\em Proc. of WWW}, ACM (2010), 621--630.

\bibitem{lotan_2012_aurora}
Lotan, G.
\newblock Big data for breaking news: Lessons from \#aurora, colorado.
\newblock Tech. rep., SocialFlow, Aug. 2012.

\bibitem{lotan_2011_news}
Lotan, G., Gaffney, D., and Meyer, C.
\newblock Audience analysis of major news accounts on twitter.
\newblock Tech. rep., SocialFlow, Aug. 2011.

\bibitem{myers_2012_external}
Myers, S.~A., Zhu, C., and Leskovec, J.
\newblock Information diffusion and external influence in networks.
\newblock In {\em Proc. of KDD}, ACM (2012), 33--41.

\bibitem{pinto_2013_predicting}
Pinto, H., Almeida, J.~M., and Gon\c{c}alves, M.~A.
\newblock Using early view patterns to predict the popularity of {YouTube}
  videos.
\newblock In {\em Proc. of WSDM}, ACM (2013), 365--374.

\bibitem{romero_2011_differences}
Romero, D.~M., Meeder, B., and Kleinberg, J.
\newblock Differences in the mechanics of information diffusion across topics:
  Idioms, political hashtags, and {ComplexContagion} on twitter.
\newblock In {\em Proc. of WWW} (2011).

\bibitem{ruan_2012_prediction}
Ruan, Y., Purohit, H., Fuhry, D., Parthasarathy, S., and Sheth, A.
\newblock Prediction of topic volume on twitter.
\newblock In {\em WebSci (short papers)} (2012).

\bibitem{szabo2012predicting}
Szabo, G., and Huberman, B.~A.
\newblock Predicting the popularity of online content.
\newblock {\em Communications of the ACM 53}, 8 (Aug. 2010), 80--88.

\bibitem{tatar_2001_predicting}
Tatar, A., Leguay, J., Antoniadis, P., Limbourg, A., de~Amorim, M.~D., and
  Fdida, S.
\newblock Predicting the popularity of online articles based on user comments.
\newblock In {\em Proc. of WIMS}, ACM (2011).

\bibitem{wu_2007_novelty}
Wu, F., and Huberman, B.~A.
\newblock Novelty and collective attention.
\newblock {\em PNAS 104}, 45 (Nov. 2007), 17599--17601.

\bibitem{yang_2011_patterns}
Yang, J., and Leskovec, J.
\newblock Patterns of temporal variation in online media.
\newblock In {\em Proc. of WSDM}, ACM (2011), 177--186.

\bibitem{yin_2012_straw}
Yin, P., Luo, P., Wang, M., and Lee, W.~C.
\newblock A straw shows which way the wind blows: ranking potentially popular
  items from early votes.
\newblock In {\em Proc. of WSDM}, ACM (2012), 623--632.

\bibitem{yu_2011_predicting}
Yu, B., Chen, M., and Kwok, L.
\newblock Toward predicting popularity of social marketing messages.
\newblock In {\em Social Computing, Behavioral-Cultural Modeling and
  Prediction}, J.~Salerno, S.~J. Yang, D.~Nau, and S.~K. Chai, Eds., vol.~6589
  of {\em LNCS}, Springer (2011), 317--324.

\end{thebibliography}
\end{document}